\documentclass{emulateapj}
\usepackage{apjfonts}

\usepackage{graphicx}

\newcommand{\La}{\mbox{${\rm Ly\alpha}$}}
\newcommand{\Line}[3]{\Ion{#1}{#2}\,$\lambda$\,#3}
\newcommand{\Lines}[3]{\Ion{#1}{#2}\,$\lambda\lambda$\,#3}
\newcommand{\Ion}[2]{#1{\,\scriptsize #2}}
\newcommand{\Mwd}{\mbox{$M_{\rm wd}$}}

\newcommand{\Msec}{\mbox{$M_{\rm sec}$}}
\newcommand{\Teff}{\mbox{$T_{\rm eff}$}}
\newcommand{\Porb}{\mbox{$P_{\rm orb}$}}

\begin{document}

\title{A 150\,MG magnetic white dwarf in the cataclysmic variable
  RX\,J1554.2+2721\altaffilmark{1}} 

\author{
B.T. G\"ansicke\altaffilmark{2},
S. Jordan\altaffilmark{3},
K. Beuermann\altaffilmark{4}, 
D. de Martino\altaffilmark{5}, \\
P. Szkody\altaffilmark{6},
T. Marsh\altaffilmark{2},
J. Thorstensen\altaffilmark{7}}

\altaffiltext{1}{Based on observations made with the NASA/ESA Hubble Space
Telescope, obtained at the Space Telescope Science Institute, which is
operated by the Association of Universities for Research in Astronomy,
Inc., under NASA contract NAS 5-26555.}
\altaffiltext{2}{Department of Physics,University of Warwick, Coventry
  CV4 7AL, UK, boris.gaensicke@warwick.ac.uk} 
\altaffiltext{3}{Institut f\"ur Astronomie und Astrophysik, Universit\"at
 T\"ubingen, Sand 1, 72076 T\"ubingen, Germany, jordan@ari.uni-heidelberg.de}
\altaffiltext{4}{Universit\"ats-Sternwarte, Geismarlandstr. 11, 37083
 G\"ottingen, Germany, beuermann@uni-sw.gwdg.de}
\altaffiltext{6}{Osservatorio di Capodimonte, Via Moiariello 16,
  I-80131 Napoli, Italy, demartin@na.astro.it}
\altaffiltext{6}{Astronomy Department, University of Washington,
  Seattle, WA 98195, USA, szkody@astro.washington.edu} 
\altaffiltext{7}{Department of Physics and Astronomy, Dartmouth
  College, 6127 Wilder Laboratory, Hanover, NH 03755-3528, USA,
  j.thorstensen@dartmouth.edu} 

\begin{abstract}
We report the detection of Zeeman-split \La\ absorption $\pi$ and
$\sigma^+$ lines in the far-ultraviolet \textit{Hubble Space
Telescope}/Space Telescope Imaging Spectrograph spectrum of the
magnetic cataclysmic variable RX\,J1554.2+2721. Fitting the STIS data
with magnetic white dwarf model spectra, we derive a field strength of
$B\simeq144$\,MG and an effective temperature of
$17\,000\,\mathrm{K}\la\Teff\la23\,000$\,K. This measurement makes
RX\,J1554.2+2721 only the third cataclysmic variable containing a
white dwarf with a field exceeding 100\,MG. Similar to the other
high-field polar AR\,UMa, RX\,J1554.2+2721 is often found in a state
of feeble mass transfer, which suggests that a considerable number
of high-field polars may still remain undiscovered.
\end{abstract}

\keywords{stars: individual (RX\,J1554.2+2721) --
          novae, cataclysmic variables          
          line: formation --
          stars: magnetic fields --
          white dwarfs
}

\section{Introduction}
In polars, a sub-class of cataclysmic variables (CV), the strong
magnetic field of the white dwarf suppresses the formation of an
accretion disk, and synchronizes the white dwarf spin with the orbital
period of the binary. Whereas single magnetic white dwarfs show a field
distribution ranging from a few tens kG to $\simeq1000$\,MG
\citep{wickramasinghe+ferrario00-1, jordan01-1, gaensickeetal02-5,
schmidtetal03-1} there has been (and still is) a discomforting lack of
CVs containing white dwarfs with magnetic fields in excess of 100\,MG.
The first high-field polar to be identified was AR\,UMa, where a field
strength of $B\simeq230$\,MG was estimated from the Zeeman-splitting
of the \La\ $\sigma^+$ component detected in a far-ultraviolet (FUV)
spectrum obtained with the \textit{IUE}
\citep{schmidtetal96-1}. Using detailed \textit{HST}/STIS data,
\citet{gaensickeetal01-1} refined the field strength measurement to
$B\simeq200$\,MG. \citet{schmidtetal01-1} reported a second high-field
polar, V884\,Her, and determined $B\simeq150$\,MG from matching a
number of narrow circularly polarized absorption features in the
optical spectrum with Zeeman-split Balmer line components.

Here, we report the identification of a third high-field polar,
RX\,J1554.2+2721, henceforth RX\,J1554. RX\,J1554 was identified as a
CV on the basis of its emission line spectrum in the Hamburg Quasar
Survey \citep{jiangetal00-1}. Its magnetic nature was suggested by
\citet{tovmassianetal01-2}, who also provided a first estimate of the
orbital period, placing the system in the $2-3$\,h period
gap. \citet{thorstensen+fenton02-1} improved the period determination
to $\Porb=151.865\pm0.009$\,min. Based on the identification of broad
bumps in their optical spectra as cyclotron harmonics, 
\citet{tovmassianetal01-2} tentatively suggested a field strength of
$B=31$\,MG for RX\,J1554.

\section{FUV spectroscopy}
\begin{figure}
\includegraphics[angle=-90,width=8cm]{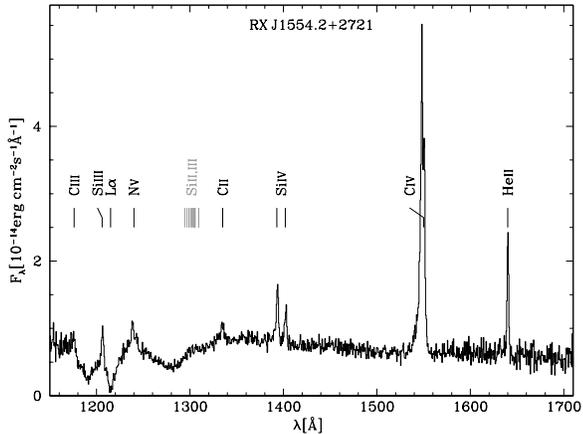}
\caption{\label{f-stis}\textit{HST}/STIS G140L snapshot spectrum of 
RX\,J1554.2+2721, along with emission line identifications.}
\end{figure}

Medium-resolution FUV spectroscopy of RX\,J1554 was obtained with
\textit{HST}/STIS on February 27, 2003, as part of our ongoing
snapshot survey of cataclysmic variables \citep{gaensickeetal03-1}. We
used the G140L grating in conjunction with the
$52\arcsec\times0.2\arcsec$ aperture, providing a spectral resolution
of $\simeq1.2$\,\AA\ and a spectral coverage of $1150-1710$\,\AA. The
STIS spectrum of RX\,J1554 (Fig.\,\ref{f-stis}) contains a blue
continuum superimposed with narrow emission lines of
\Line{C}{III}{1176}, \Line{Si}{III}{1206}, \Line{N}{V}{1240},
\Line{C}{II}{1335}, \Lines{Si}{IV}{1393,1402}, \Line{C}{IV}{1550}, and
\Line{He}{II}{1640}. A weak broad emission bump centered on
$\sim1300$\,\AA\ may be related to the \Lines{Si}{III}{1294--1309}
multiplet, the \Lines{Si}{II}{1260,65} doublet is not detected. A most
unusual feature is the broad ($\sim75$\,\AA) absorption line centered
on $\sim1280$\,\AA. It falls clearly blue-ward of the 1400\,\AA\
quasi-molecular $H_2^+$ absorption observed in cool white
dwarfs. Considering the magnetic nature of RX\,J1554, a tantalising
possible identification for this feature is the $\sigma^+$ component
of \La\ Zeeman-split in a field $B>100$\,MG, analogous to that
observed in the FUV spectrum of AR\,UMa \citep{schmidtetal96-1,
gaensickeetal01-1}. We measured a F28$\times$50LP magnitude of
$\simeq16.8$ (roughly equivalent to the $R$ band) from the STIS CCD
acquisition image taken prior to the FUV spectroscopy. The system was
hence in a state of relatively low accretion activity
\citep{tovmassianetal01-2, thorstensen+fenton02-1}.

\section{Spectral Analysis}
We have analysed the STIS spectrum of RX\,J1554 using magnetic white
dwarf model spectra computed with the code of \citet{jordan92-1} and a
fitting routine very similar to the one described by
\citet{euchneretal02-1}. In brief, our fit involves three free
parameters: the magnetic field strength $B$ (where we restricted our
models to centered dipole fields), the angle between the line-of-sight
and the magnetic axis $\Psi$, and the effective temperature of the
white dwarf \Teff. Flux and circular polarization spectra are
calculated for a large number of surface elements representing the
surface of the white dwarf, and the actual white dwarf model spectrum
is obtained by adding up the contributions of all visible elements. In
contrast to \citet{euchneretal02-1}, we used 12 viewing angles
(instead of 9), and interpolated in a grid of 12 propagation angles
within the atmosphere to account for the effect of limb darkening
(instead of a linear approximation). The optimization is initially
done using a genetic algorithm \citep{carroll01-1} and subsequently
refined by a downhill simplex algorithm. Once the best-fit parameters
are determined, a final model is calculated by consistently solving
the radiative transfer equations \citep{jordan92-1} in order to avoid
the interpolations used in the fitting code. 

Whereas the position of the \La\ Zeeman components as well as their
profiles allow a relatively accurate measurement of $B$ and $\Psi$,
determining the effective temperature of high-field magnetic white
dwarfs is notoriously difficult. In non-magnetic white dwarfs, the
analysis of the Stark-broadened Balmer and/or Lyman lines is most
useful in establishing the effective temperature. However, the
combination of Stark broadening and Zeeman splitting at high field
strengths is still too poorly understood to use the hydrogen line
profiles as exact temperature indicators in strongly magnetic white
dwarfs. Similarly, also the slope of the continuum is subject to
uncertainties in the atomic physics. A general trend at high fields
appears to be that the optical continuum implies higher temperatures
than the FUV continuum, as observed e.g. in PG\,1031+234
\citep{schmidtetal86-1} and in AR\,UMa \citep{gaensickeetal01-1}.

Our best-fit to the STIS spectrum of RX\,J1554 results in
$B=144\pm15$\,MG, $\Teff\sim17500$\,K and
$\Psi=38\pm15^{\circ}$. Thus, RX\,J1554 is the third CV containing a
magnetic white dwarf with a field exceeding 100\,MG. The field
strength determined here from the unambigous detection of the \La\
Zeeman $\sigma^+$ component exceeds the estimated $B=31$\,MG of
\citet{tovmassianetal01-2} by a large factor. The most likely cause
for this disagreement is an erroneous assignment of cyclotron harmonic
numbers to the weak emission bumps observed in the blue part of the
optical spectrum \citep{tovmassianetal01-2}. For a field strength of
$\sim150$\,MG, the emission of the cyclotron fundamental is centered
at $\sim7200$\,\AA, the first harmonic at $\sim3600$\,\AA. Neither the
spectra of \citet{tovmassianetal01-2} nor those of
\citet{thorstensen+fenton02-1} display any feature readily
identifiable as the cyclotron fundamental or first
harmonic. Phase-resolved spectropolarimetry would be desirable to
probe the nature of the optical continuum emission.

\begin{figure}
\includegraphics[angle=-90,width=8cm]{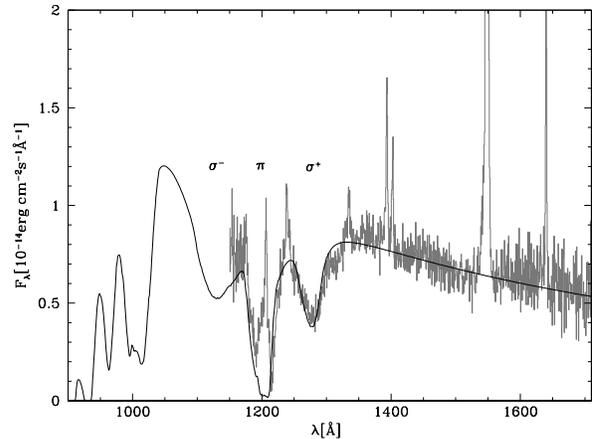}
\caption{\label{f-stisfit}The best-fit magnetic white dwarf model
  spectrum ($B=144$\,MG, $\Teff=17\,500$\,K) reproduces well the
  position and profile of the observe \La\ $\pi$  and $\sigma^+$
  components. }
\end{figure}

Figure\,\ref{f-stisfit} shows the STIS spectrum along with
our best-fit model extended blue-wards up to the Lyman limit. It is
apparent that future phase-resolved \textit{FUSE} and \textit{HST}
spectroscopy of RX\,J1554 would be extremely useful to confine in more
detail the topology of the magnetic field in RX\,J1554, which could
deviate from a simple dipole geometry (as shown in the case of AR\,UMa
by \citealt{gaensickeetal01-1} and
\citealt{ferrarioetal03-1}). \textit{FUSE} observations would also
provide an additional constraint on the effective temperature of the
white dwarf.

\begin{figure}
\includegraphics[angle=-90,width=8cm]{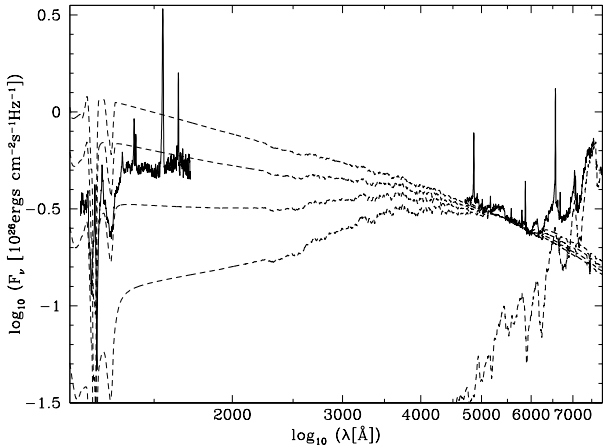}
\caption{\label{f-sed} Combined ultraviolet/optical spectrum of
  RX\,J1554, using our STIS data and the mean low-state spectrum of
  \citet{thorstensen+fenton02-1}. Overplotted as dashed lines are
  (from top to bottom) magnetic white dwarf models ($B=144$\,MG,
  $\Psi=28^{\circ}$ with \Teff\,=\,30\,000\,K, 25\,000\,K, 20\,000\,K,
  and 15\,000\,K. The models have been scaled to the $V$-band flux of
  RX\,J1554, after subtracting the contribution from the companion
  star.}
\end{figure}

In Fig.\,\ref{f-sed} we show an average low-state spectrum of
RX\,J1554 from \citet{thorstensen+fenton02-1} along with our STIS
data. The red part of the optical spectrum clearly reveals the
secondary star, the blue continuum is most likely associated with
photospheric emission from the white dwarf. Following
\citet{thorstensen+fenton02-1}, the spectral contribution of the
secondary star is well described by a M4 template scaled for a
distance of $d=210$\,pc and a Roche-lobe equivalent radius of the
secondary of $R_{L_2}=1.76\times10^{10}$\,cm (corresponding to a mass
ratio $\Mwd/\Msec=3$). Whereas the relatively flat spectral slope of
the STIS spectrum is best described with a temperature of
$\simeq17\,500$\,K, the combined optical-FUV spectral energy
distribution suggests a higher temperature of $\sim23\,000$\,K
(reddening is negligible along the line of sight to RX\,J1554). At a
distance of $d=210$\,pc, the flux scaling factors implied by the
models shown in Fig.\,\ref{f-sed} give white dwarf radii of
$9.8\times10^8$\,cm, $7.9\times10^8$\,cm, and $6.6\times10^8$\,cm for
$\Teff=15\,000$\,K, 20\,000\,K, and 25\,000\,K~--~which are all
plausible, but do not help to constrain the temperature of the white
dwarf. We conclude that the temperature of the white dwarf in
RX\,J1554 is likely to be $17\,000\,\mathrm{K}\la\Teff\la23\,000$\,K.

\section{Magnetic fields in white dwarfs}
\begin{figure}
\includegraphics[width=8cm]{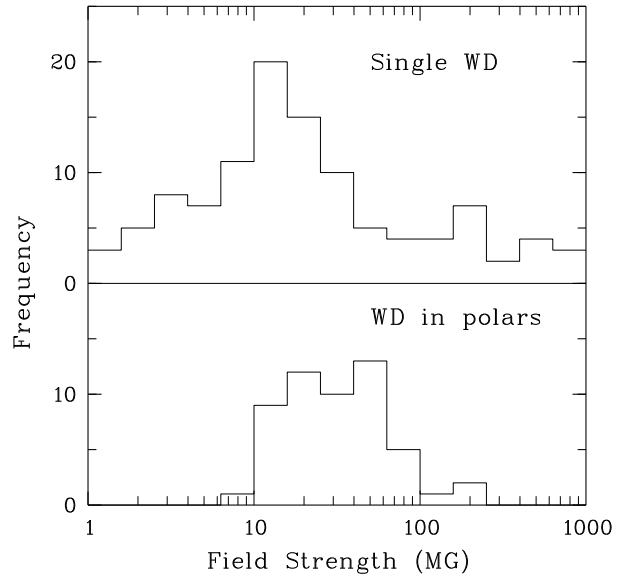}
\caption{\label{f-fielddist} Distribution of magnetic field strength
  for single white dwarfs and for white dwarfs in CVs.}
\end{figure}

The statistics of magnetic fields in single white dwarfs has been
substantially improved by the first results from the Sloan Digital Sky
survey \citep{gaensickeetal02-5, schmidtetal03-1}, even though
\citet{liebertetal03-1} argue that the fraction of magnetic white
dwarfs may still be significantly underestimated. Combining the SDSS
results with those summarized in \citet{wickramasinghe+ferrario00-1},
the number of single white dwarfs with known field strength is 108. Our
current census of polars is 76, of which some measure of the surface
field is available for a subset of 53. Figure\,\ref{f-fielddist} shows
the distribution of the measured (mean or typical) field strengths in
the single white dwarfs and in polars. For the latter, the quoted
field strength is preferentially that of the main accreting pole as
determined from the spacing of cyclotron lines, replaced by Zeeman
detections where no cyclotron measurement is available. It is likely
that this distribution is biased against the detection of both high
and low field strengths, because the high-field systems have weak
spectral features in the optical range and the low-field systems often
display a cyclotron continuum with all lines washed out. We suspect
that most of the 23 polars with no field measurement and all of the 40
confirmed or suspected intermediate polars in the 2003 edition of the
\citeauthor{ritter+kolb03-1} catalogue have field strengths typically in
the 1--25\,MG range, given that the minimum field strength which
allows a magnetic CV to synchronize is of the order of 3\,--\,10\,MG.

\section{Discussion}
The lack of high-field polars has been a long-standing puzzle.
\citet{hameuryetal89-3} suggested that for strong fields ($B>40$\,MG)
the wind of the secondary may couple to the magnetic field lines of
the white dwarf rather to those of the secondary, and, as a
consequence of the large magnetospheric radius of the white dwarf,
magnetic braking would be much more efficient than in non-magnetic
CVs~--~in fact so efficient that the mass transfer would be thermally
unstable, with a correspondingly very short life time of such systems. 

\citet{wickramasinghe+wu94-1} and \citet{lietal94-1} alternatively
argued that in high-field polars the magnetic field lines of the
secondary will form closed loops or connect to the field lines of the
white dwarf, reducing the magnetic flux in open field
lines. Therefore, their prediction is that the angular momentum loss
and mass transfer rates in strongly magnetic CVs should be lower
compared to non-magnetic ones, and they predict a maximum field
strength of $\sim70-100$\,MG, above which mass transfer will be
suppressed completely (see, however, the discussion by
\citealt{kingetal94-1}).

Our determination of a white dwarf magnetic field strength of
$B\simeq150$\,MG in RX\,J1554 raises the number of high-field polars
($B>100$\,MG) to three~--~which makes these objects a rare but no
longer exceptional species. AR\,UMa ($B\simeq200$\,MG,
$\Porb=116$\,min) and V884\,Her ($B\simeq150$\,MG, $\Porb=113$\,min)
are both below the period gap, and RX\,J1554 ($B\simeq145$\,MG,
$\Porb=152$\,min) is right in the middle of the gap~--~i.e. all known
high-field polars have periods below 3\,h. It seems, however,
premature to draw any conclusion based on three systems only.  One
property that at least AR\,UMa and RX\,J1554 share is that they are
frequently encountered in states of low mass transfer
\citep{remillardetal94-2, schmidtetal96-1, schmidtetal99-1,
tovmassianetal01-2, thorstensen+fenton02-1}, while the long-term
variability of V884\,Her is less well documented. Assuming that
frequent low states are a general property of high-field polars it is
likely that a considerable fraction of these systems have evaded
discovery so far. The exact causes of low states in polars are
not yet fully understood, possible mechanisms that could decrease the
mass transfer rate on time scales of years include starspots moving
across the $L_1$ point \citep{king+cannizzo98-1, hessmanetal00-1} and
the swinging dipole model by \citet{andronov87-1}. Detailed long-term
monitoring of the accretion activity of polars in general, and
high-field polars in particular, would be desirable to establish a
measure of their accretion rates following the approach of
\citet{hessmanetal00-1}.

\citet{schmidtetal03-1} believe that the maximum in the distribution
of single white dwarfs near 20\,MG is real. Taking into account the
selection effects for identifying magnetic CVs in the first place, and
measuring the magnetic fields of their white dwarfs in the second
place, it seems possible that the intrinsic distribution of magnetic
field strengths in CV white dwarfs is not dissimilar from the one of
single white dwarfs. RXJ\,1554 is one example of the systems not
readily recognisable as high-field polars by optical means, and
systems with white dwarfs of still higher field strength possibly may
exist but appear no different than ordinary polar.

BTG was supported by a PPARC Advanced Fellowship.  Additional support
was provided through NASA grant GO-9357 from the Space Telescope
Science Institute, which is operated by the Association of
Universities for Research in Astronomy, Inc., under NASA contract NAS
5-26555.

\bibliographystyle{apj}
\bibliography{aamnem99,aabib}

\end{document}